\documentclass[prl,aps,showpacs,twocolumn,superscriptaddress]{revtex4}

\usepackage{amsmath}
\usepackage{amssymb}
\usepackage{bm}      
\usepackage{epsfig}
\usepackage{graphicx}

\def\ra{\rangle}
\def\la{\langle}

\newcommand{\ket}[1]{ {\left| #1 \right\rangle} }

\begin{document}
\title{A symmetry analyzer for non-destructive Bell state detection using EIT}

\author{S. D.\ Barrett} \email{sean.barrett@hp.com}
\affiliation{Hewlett-Packard Laboratories, Filton Road Stoke Gifford,
  Bristol BS34 8QZ, UK}

\author{Pieter Kok} 
\affiliation{Hewlett-Packard Laboratories, Filton Road Stoke Gifford,
  Bristol BS34 8QZ, UK}

\author{Kae Nemoto} 
\affiliation{National Institute of Informatics, 2-1-2 Hitotsubashi,
  Chiyoda-ku, Tokyo 101-8430, Japan}

\author{R. G.\ Beausoleil} 
\affiliation{Hewlett-Packard Laboratories, 13837 $175^{th}$ Pl.\ NE,
  Redmond, WA 98052-2180, USA}

\author{W. J.\ Munro} 
\affiliation{National Institute of Informatics, 2-1-2 Hitotsubashi,
  Chiyoda-ku, Tokyo 101-8430, Japan}
\affiliation{Hewlett-Packard Laboratories, Filton Road Stoke Gifford,
  Bristol BS34 8QZ, UK}

\author{T. P.\ Spiller} 
\affiliation{Hewlett-Packard Laboratories, Filton Road Stoke Gifford,
  Bristol BS34 8QZ, UK}

\date{\today}

\begin{abstract}
 We describe a method to project photonic two-qubit states onto
 the symmetric and antisymmetric subspaces of their Hilbert space.
 This device utilizes an ancillary coherent state, together with a
 weak cross-Kerr non-linearity, generated, for example, by
 electromagnetically induced transparency. The symmetry analyzer
 is non-destructive, and works for small values of the cross-Kerr
 coupling. Furthermore, this device can be used to construct a
 non-destructive Bell state detector.
\end{abstract}


\maketitle

\noindent Two-qubit measurements are an important resource in
Quantum Information Processing (QIP), enabling key applications
such as the teleportation of states and gate, dense coding and error correction. 
In particular, a measurement device that does
not destroy the qubits is a very powerful tool, since it allows
entanglement distillation \cite{bennett96} and efficient quantum computing based on
measurements \cite{nielsen,leung,raussendorf}. This is especially
useful when the qubits interact weakly, and interaction-based quantum
gates are hard to implement (for example, photonic qubits have negligible
interaction). Furthermore, a non-destructive two-qubit measurement
device can act as an deterministic source of entangled qubits.

Optical QIP is of special interest, because electromagnetic fields are
ideal information carriers for long distance quantum
communication. Photonic quantum states generally suffer low
decoherence rates compared to most massive qubit systems, but we need
optical information processing devices that overcome the negligible
interaction between the photons. Optical quantum computation and
communication will therefore benefit greatly from non-destructive
two-qubit measurements. Arguably the most important two-photon
measurement is the measurement in the maximally entangled Bell
basis. When the computational basis of a single-photon qubit is given
by two orthogonal polarization states ($H$ and $V$), then the Bell
states can be written as $|\Psi^{\pm}\ra = (|H,V\ra \pm
|V,H\ra)/\sqrt{2}$ and $|\Phi^{\pm}\ra = (|H,H\ra \pm
|V,V\ra)/\sqrt{2}$. A non-destructive Bell measurement then projects
the two photons onto one of the Bell states. This can be used in the
teleportation of probabilistic gates into optical circuits
\cite{gottesman99,KLM01}, and consequently enables efficient linear optical
quantum computing. In addition, a deterministic non-destructive Bell
measurement would also act as a bright source of entangled photons.

Braunstein and Mann presented a linear optical method to
distinguish two out of the four optical Bell states
\cite{braunstein95}. In 1999, it was shown independently by
Vaidman and Yoran, and L\" utkenhaus {\em et al.} that the
Braunstein-Mann method is optimal \cite{yoran,Lutkenhaus99}: When
one is restricted to linear optics and photon counting (including
feed-forward processing) at most half of the Bell states can be
identified perfectly. This detection method is therefore {\em
probabilistic}. Furthermore, it destroys the photons in the photon
counting process, and is thus of limited use in efficient
large-scale QIP.

One way to improve on this scheme is to move beyond linear optics,
i.e. to induce an interaction between the photons. This can be
achieved using a cross-Kerr medium, i.e., a nonlinear medium
that can be described by an interaction Hamiltonian of the form
\begin{equation}\label{crossKerreq}
  \hat H_K=\hbar \chi\, \hat{n}_a \hat{n}_c\, ,
\end{equation}
where $\hat{n}_k$ is the number operator for mode $k$, and
$\hbar\chi$ is the coupling strength of the nonlinearity. A photon
in mode $c$ will then accumulate a phase shift $\theta = \chi t$
that is proportional to the number of photons in mode $a$. Such a
medium can be used as an optical switch \cite{duguay}. More to the
point, when the nonlinearity is large (i.e., $\theta \approx
\pi$), it naturally implements a controlled-phase gate at the
single photon level. This inspired applications such as photon
number quantum non-demolition (QND) measurements
\cite{milburn84,imoto85}, Noon-state
generation \cite{gerry}, a Fredkin gate \cite{milburn89},
and culminated in a full-scale proposal for optical quantum
computers \cite{dariano}. In particular, with a large nonlinearity
we can build a Bell state analyzer \cite{martin,cavitynote}.

Unfortunately, natural Kerr media have extremely small
nonlinearities, with a typical dimensionless magnitude of $\theta
\approx 10^{-18}$ \cite{boyd,kok}. A large Kerr  nonlinearity at
the single-photon level is therefore practically impossible.
However, there are ways to make nonlinearities of  magnitude
$\sim10^{-2}$, for example with Electromagnetically Induced
Transparencies (EIT) \cite{harr99,munro03,harris04},
whispering-gallery micro-resonators \cite{kipp04}, optical fibers
\cite{li04}, or cavity QED systems \cite{grang98,kimble95}. In
this Letter, we show how to build a nondestructive interferometric
Bell state analyzer with such small-but-not-tiny Kerr
nonlinearities, and additional coherent state resources.

As a specific example of a very promising method
for generating the form of non-linearity required, we
consider EIT in condensed matter systems. We
have analyzed a model system at length \cite{munro03}, 
considering three photon modes
interacting through dipole couplings to a four-level $\mathcal{N}$
atomic system \cite{beau03}. Mode $a$ generally describes a Fock
state $\ket{n_a}$, and mode $c$ is a coherent state
$\ket{\alpha_c}$; these two fields interact through a third pump
mode that is a sufficiently intense coherent state that both it
and the internal atomic energy levels can be factored out of the
evolution, creating an effective non-linear Kerr interaction
between modes $a$ and $c$ of the form given in Eq. (\ref{crossKerreq}). 
In general, it is difficult to achieve
a substantial vacuum Rabi frequency using free-space
fields \cite{vane00}, but encapsulating one or more atoms in a
waveguide (such as a line defect in a photonic crystal structure)
allows field transversality to be maintained at mode
cross-sectional areas that have dimensions smaller than the
optical wavelengths of the interacting fields. A two-dimensional
photonic crystal waveguide constructed from diamond thin film with
nitrogen-vacancy color centers fabricated in the center of the
waveguide channel \cite{hemm01,shah02} could provide a sufficiently
large nonlinearity to realize our method experimentally. For
example, a cryogenic NV-diamond system with $2 \times 10^4$ color
centers can generate a phase shift of more than 0.1~radians per
signal photon with a probe photon number $n_c = \alpha_c^2 = 1.3
\times 10^4$ and modest detunings.

%
%
We turn now to the application of such non-linearities for
Bell state analysis.
As mentioned above, it is well known that a beam-splitter can be
used to discriminate between the singlet and the remaining triplet
Bell states \cite{braunstein95}. If the two incoming modes are
combined on a beam splitter, the Bell states are transformed as
\begin{eqnarray}
|\Psi^{-}\ra = |H,V\ra - |V,H\ra &\rightarrow& |H,V\ra - |V,H\ra ,\nonumber \\
|\Psi^{+}\ra = |H,V\ra + |V,H\ra &\rightarrow& |HV,0\ra - |0,HV\ra , \nonumber \\
|\Phi^{\pm}\ra = |H,H\ra \pm |V,V\ra &\rightarrow&  |H^2,0\ra -|0,H^2\ra  \nonumber \\
&&\;\;\;\pm |V^2,0\ra \mp |0,V^2\ra . \label{BeamsSplitterEq}
\end{eqnarray}

After the beam-splitter transformation, the singlet state,
$|\Psi^{-}\ra$, is \emph{balanced}, i.e. it has only one photon in
each spatial mode. On the other hand, the triplet states are
\emph{bunched}, i.e. they have coherent superpositions of either zero
or two photons in each spatial mode. Our scheme proceeds by
non-destructively distinguishing between these two cases, and
subsequently transforming the states back to the Bell basis using
a second beam splitter. This non-destructive symmetry analysis
therefore allows the singlet state to be discriminated from the
triplet states. As we discuss further below, a full
non-destructive Bell measurement can be implemented by repeated
applications of the symmetry analysis, interleaved with
appropriate local operations.

It is important to note that the balanced and bunched states must
be discriminated in such a way that no other information is
discovered about the states. In particular, determining the
number of photons in a particular spatial mode, even
non-destructively, would destroy the coherence of the bunched
states. For this reason, existing photon number QND measurement
techniques \cite{milburn84,imoto85,munro03} with small $\theta$ 
are insufficient to perform the symmetry analysis step. The technique
for non-destructive symmetry analysis that we describe below is one
of the principal results of this work.

\begin{figure}[!hbt]
\begin{center}
\includegraphics[scale=1]{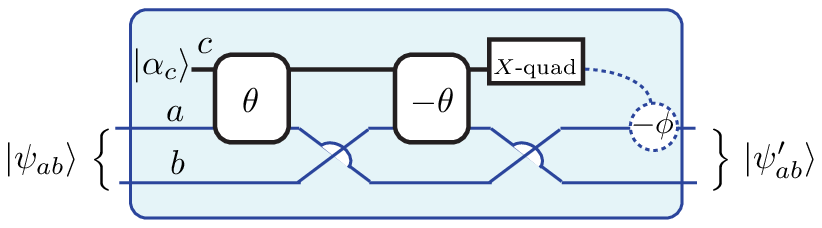}
\end{center}
\caption{Schematic circuit showing how two relatively small
cross-Kerr non-linearities can be used to discriminate between
$(|2,0\ra \pm |0,2\ra)/\sqrt{2}$ and $|1,1\ra$ non-destructively.
The boxes labeled by $\theta$ and $-\theta$  represent the
cross-Kerr non-linearities that induce a phase shift in the
coherent state $|\alpha_c\ra$ proportional to the number of
photons in the corresponding signal mode. A homodyne measurement
of the $X$-quadrature with outcome $x$ will then discriminate
between the two input states. Depending on this outcome, a
relative phase shift $2 \phi(x)$ is needed to restore the $|2,0\ra
\pm  |0,2\ra$ state.} \label{fig2}
\end{figure}

%
%
In order to describe our scheme for symmetry analysis, we first
consider the illustrative example of an analyzer capable of
non-destructively distinguishing between the balanced and bunched
states, $|1,1\ra$ and $(|2,0\ra \pm |0,2\ra)/\sqrt{2}$. Here,
$|j,k\ra$ represents a two spatial mode optical state, with $j$
photons in mode $a$ and $k$ photons in mode $b$, with all photons
polarized in the same direction. The setup for discriminating
these states is shown in Fig.~\ref{fig2}. Mode $c$ is initially
prepared in a coherent state $|\alpha_c\ra = e^{-|\alpha_c|^2/2}
\sum_n \alpha_c^n/\sqrt{n!} | n \ra$, where $| n \ra$ denotes an
$n$-photon number state \cite{WallsMilburnBook}. The coherent
state can be generated, for example, by a laser pulse. The photons
in mode $c$ sequentially interact with those in modes $a$ and $b$
via the two (relatively small) cross-Kerr non-linear operations,
acting with phases $\theta$ and $-\theta$ respectively. These
operations can be written as $\exp(i\theta \hat{n}_a \hat{n}_c)$
and $\exp(-i\theta \hat{n}_b \hat{n}_c)$ as follows from 
Eqn (\ref{crossKerreq}) \cite{note}.

%
%

\begin{figure}[!hbt]
\begin{center}
\includegraphics[scale=1]{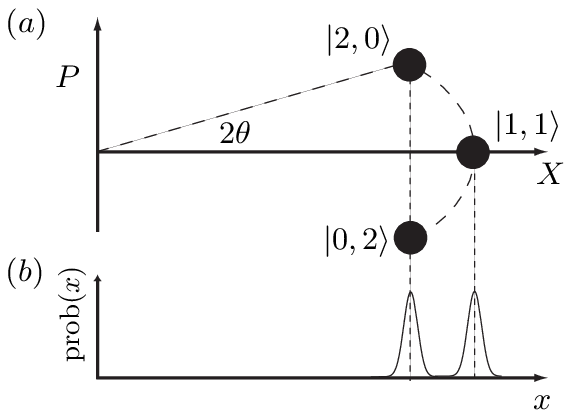}
\end{center}
\caption{(a) Schematic phase space illustration of the state $|\psi_1\ra =
d_1 |1,1\ra |\alpha_c\ra + d_2 \frac{1}{\sqrt{2}} ( |2,0\ra
|\alpha_c e^{2i\theta}\ra \pm |0,2\ra |\alpha_c e^{-2i\theta}\ra$.
For the $|1,1\ra$ component of the state the, the probe mode
receives zero total phase shift, whereas for the $|2,0\ra$ and
$|0,2\ra$ components, the phase shifts are $+2 \theta$ and
$-2\theta$, respectively. (b) Corresponding probability
distribution for the outcome of the X-quadrature measurement of the
probe beam. The
two peaks in this distribution are associated with the states
$|1,1\ra$ and $|2,0\ra \pm |2,0\ra$.} \label{figPhaseSpaceCartoon}
\end{figure}

%
%
%
Suppose now that the input state for the two signal modes $a$ and
$b$ and the probe mode $c$ is given by
\begin{equation}
  |\psi_0\ra = \left[ d_1 |1,1\ra +  \frac{d_2}{\sqrt{2}}\left(
  |2,0\ra \pm |0,2\ra \right) \right] |\alpha_c\ra \, ,
\end{equation}
where $d_1$, $d_2$ are complex coefficients satisfying the usual
normalization requirements. The effect of each cross-Kerr
operation is to induce a phase shift in the coherent state
$|\alpha_c\ra$, which is proportional to the number of photons in
the corresponding signal mode (alternatively although each Fock 
state in the coherent state imparts a different Kerr shift to the 
signal photons, the measurement of the coherent state is designed 
so that there is only one overall phase shift on the signal 
photons afterwards). Thus, for the $|1,1\ra$ component
of the state, the total phase shift induced is $\theta + (-\theta)
= 0$, and for the $|2,0\ra$ and $|0,2\ra$ components, the phase
shifts are $+2 \theta$ and $-2\theta$, respectively. After these
cross-Kerr operations, the state of the three modes is thus given
by
\begin{equation}
|\psi_1\ra = d_1 |1,1\ra |\alpha_c\ra + \frac{d_2}{\sqrt{2}} (
|2,0\ra |\alpha_c e^{2i\theta}\ra \pm |0,2\ra |\alpha_c
e^{-2i\theta}\ra ).
\end{equation}
This state is illustrated in the phase space plot in Fig.
\ref{figPhaseSpaceCartoon} (a).

In order to distinguish the balanced and bunched components of
$|\psi_1\ra$, it is sufficient to measure the $X$-quadrature
component of the probe mode $c$. This can be achieved with a
standard homodyne measurement. To perform such a measurement, the
probe mode is combined at a beam splitter with a local oscillator of
the same frequency. The output is then measured with photo-detectors
\cite{WallsMilburnBook}. Provided the local oscillator is
prepared in a large-amplitude coherent state with the same phase
as $|\alpha_c\ra$, homodyne measurement amounts to a (destructive)
measurement of the observable $\hat{X} = \hat{c} + \hat{c}^\dag$,
where $\hat{c}$ is an annihilation operator for photons in the probe
mode. Using the result $\la x|\beta\ra = (2\pi)^{-1/4} \exp[
-\mathrm{Im}(\beta)^2 - (x-2\beta)^2/4]$ \cite{GardinerZollerBook}, where $|x\ra$ is
an eigenstate of $X$ with eigenvalue $x$, the state of modes $a$ and $b$ after
the measurement on $c$ is
\begin{eqnarray}\label{measuredstate}
 |\psi_3\ra &=&d_1 f(x,\alpha_c) |1,1\ra +\\ && \frac{d_2}{\sqrt{2}}  f(x,\alpha_c\cos
 2\theta) \left( e^{i\phi(x)} |2,0\ra \pm e^{-i\phi(x)} |0,2\ra
 \right), \nonumber
\end{eqnarray}
where we have defined
\begin{eqnarray}
 f(x,\beta) &\equiv& (2\pi)^{-1/4}\exp \left[-(x-2\beta)^2/4
 \right]\cr \phi(x) &\equiv& \alpha_c \sin 2\theta (x - 2\alpha_c\cos
 2\theta) \mod 2\pi\, .
\end{eqnarray}
The Gaussian terms $f(x,\alpha_c)$ and $f(x,\alpha_c\cos 2\theta)$
in (\ref{measuredstate}) correspond to probability amplitudes
associated with each of the two states $|1,1\ra$ and $|2,0\ra \pm
|2,0\ra$ respectively [see Fig. \ref{figPhaseSpaceCartoon}(b)].
The phase shift $\phi(x)$ associated with the two-photon
components depends on the outcome of the homodyne measurement.
This can be corrected by applying the phase shift operation
$\exp[-i \phi(x) \hat{n}_a ]$, conditional on the obtained value
of $x$. In order to resolve the balanced and the bunched
components, we require only a small overlap between their
probability distributions. Values of $x$ below the mid-point 
between the peaks define one measurement outcome, and values of $x$ 
above it the other outcome. The error probability is thus the sum of the 
lower $x$ distribution tail above the midpoint and the upper 
distribution tail below the mid-point, and is given by
$P_\text{error} = \text{erfc}(\sqrt{2}\alpha_c\theta^2)/2$. 
This is less than $0.01$ provided  $\alpha_c\theta^2 > 1.2$. Highly
accurate discrimination is therefore possible with weak cross-Kerr
non-linearities ($\theta \ll \pi$) provided $\alpha_c$ can be made
sufficiently large. For example, the system described in the
NV-diamond example given above generates the error probability
$P_\text{error} = 0.01$.

\begin{figure}[!htb]
\begin{center}
\includegraphics[scale=0.9]{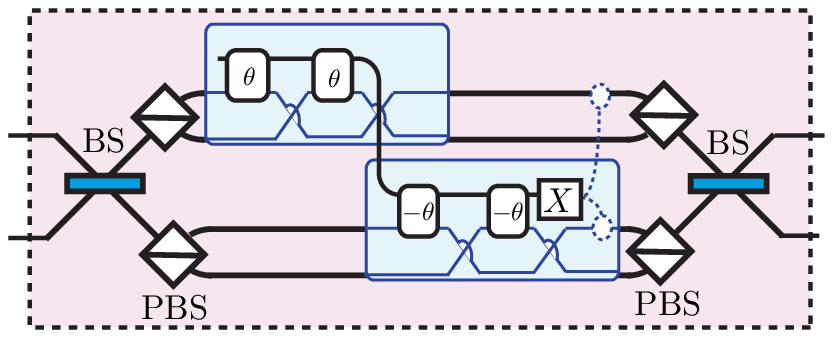}
\end{center}
\caption{The symmetry analyzer: photonic two-qubit input states
 interfere on a 50:50 beam splitter (BS). The polarizing beam
 splitters (PBS) will transform the beam splitter output into four
  modes with equal polarization. Using the methodology of
  Fig.~\ref{fig2}, one can distinguish between the resulting bunched and
  balanced states. This results in a measurement of
  the symmetry of the photonic two-qubit input state. The PBS and BS
  after the homodyne measurement return the two modes to the two-qubit
  space.}
  \label{fig3}
\end{figure}

A straightforward generalization of the methodology described
above can be used to construct a non-destructive photonic symmetry
analyzer on the two-qubit Hilbert space $\mathcal{H}$ of the
incoming modes, spanned by the states $|H,H\ra$, $|H,V\ra$,
$|V,H\ra$, and $|V,V\ra$ (see Fig.~\ref{fig3}). As noted above,
the beam splitter transformation, Eq. (\ref{BeamsSplitterEq}),
transforms the singlet state into a balanced state, whereas the
triplet states are bunched. The polarizing beam splitters (PBS)
will separate the two polarization modes. A polarization rotation
(not shown) applied to the same output of each PBS will then
ensure that all the photons have identical polarization, thus
satisfying the assumption made about the inputs to the analyzer of
Fig.~\ref{fig2}. (These rotations are then undone before the
outgoing PBS.) By counting the phase shifts ($\theta$ and
$-\theta$) we can determine the total phase that is acquired by
$|\alpha_c\ra$. As before, the (balanced) singlet state will not
induce a phase shift in $|\alpha_c\ra$. The different components
of the (bunched) triplet states will induce phase shifts of
$+2\theta$ or $-2\theta$. Therefore, $X$-quadrature homodyne
measurement of mode $c$ now allows the singlet and triplet states
to be distinguished non-destructively. After corrective phase
shifts (again conditional on the outcome of the $X$-quadrature
measurement) and recombination on the PBS, the final beam splitter
will return the state to the two-qubit Hilbert space
$\mathcal{H}$. Note that it is crucial that the measurement does
not introduce decoherence between the symmetric amplitudes, as
repetition of the symmetry analysis is needed for full Bell state
analysis.

\begin{figure}[!htb]
\begin{center}
\includegraphics[scale=1]{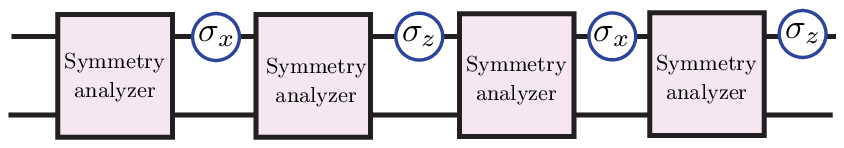}
\end{center}
\caption{The Bell-state detector: repeated application of the
symmetry analyzer (SA) and local unitary rotations in $\mathcal{H}$
subsequently project onto different Bell states. The local unitaries are
simply performed with suitable optical plates for polarization qubits.}
\label{fig4}
\end{figure}

Once we have a non-destructive symmetry analyzer (SA), it is
straightforward to construct a quantum non-demolition Bell-state
detector (depicted in Fig.~\ref{fig4}). First, we test whether or
not the input state is the singlet by applying the SA. We then
apply a bit flip $\sigma_x$ to move the antisymmetric subspace
into the symmetric subspace. We apply the SA again, and if the
transformed input state is the singlet, we know that the original
state was $|\Phi^-\ra$. We then apply a relative phase shift
$\sigma_z$. If the third SA finds the singlet, then the input was
$|\Phi^+\ra$. If no singlet signal has arisen, then the input
state must have been $|\Psi^+\ra$. We can test this by applying
another bit flip $\sigma_x$ and invoke the SA again. The final
$\sigma_z$ ensures that the outgoing state is actually that
identified by the analyzer. Clearly if one is prepared to accept
not finding the singlet in the first three analyzers as the
signature of $|\Psi^+\ra$, the final analyzer can be omitted. In
this case the third and final single-qubit operation is instead
$\sigma_y$ to restore the outgoing state to that identified.
Furthermore, if classical switching conditional on the homodyne
measurement results is employed, the analysis could be terminated
after a singlet signal from any SA, by switching the two-qubit
state out and then reconstructing the identified Bell state by a
local operation.

To summarize, in this letter we have shown how to construct a Bell
state analyzer from small cross-Kerr non-linearities---small here
means much less than the size of non-linearity required to perform
a maximally entangling/disentangling gate directly between
photons. Our analyzer distinguishes all four polarization Bell
states and is near deterministic in operation. We have suggested
EIT systems as one potential route for realizing the required
cross-Kerr non-linearities, which could lead to practical QIP in
the relatively near future, especially since for our proposed
symmetry and Bell state analyzers there is not a requirement to
generate $\pi$ phase shifts. As we have shown, as long as the
probe beam has a sufficient amplitude $\alpha_c$ such that
$\alpha_c \theta^2 \gtrsim 1$, we can work with much smaller phase
shifts. This makes our analyzers rather easier to implement that
those based on standard non-linear quantum logic.

\noindent
{\em Acknowledgments}: This work was supported by JSPS and 
the European Union Nanomagiqc and Ramboq projects. 
KN was supported in part by MPHPT and Asahi-Glass
research grants.

\bibliography{bsa}

\begin{thebibliography}{33}
\bibitem[{cav()}]{cavitynote}
\bibinfo{note}{The strong cross-Kerr non-linearity is achieved using a small
  non-linearity in a cavity, and operating in the strong coupling regime.}
\bibitem[{not()}]{note}
\bibinfo{note}{The arrangement of non-linearities is not unique. For example, a
  non-linearity (of strength $2\theta$ for the same discrimination) could be
  applied to just one mode, and the quadrature $c^{\dagger} e^{-2i\theta} + c
  e^{2i\theta}$ for probe mode $c$ measured.}
\end{thebibliography}

\end{document}